\documentstyle[titlepage,12pt]{article}
\def\strut{\rule[-.5cm]{0cm}{1cm}}

\def\dspace{\baselineskip = .30in}

\long\def\@makefntext#1{\parindent 0cm\noindent
\hbox to 1em{\hss$^{\@thefnmark}$}#1}

\begin{document}
\newfont{\bg}{cmr10 scaled\magstep3}
%
\newcommand{\gsimeq}
{\hbox{ \raise3pt\hbox to 0pt{$>$}\raise-3pt\hbox{$\sim$} }}
\newcommand{\lsimeq}
{\hbox{ \raise3pt\hbox to 0pt{$<$}\raise-3pt\hbox{$\sim$} }}


\newcommand{\pletb}[3]{Phys. Lett. {\bf B#1}, (#2) #3}
\newcommand{\prevlet}[3]{Phys. Rev. Lett. {\bf #1}, (#2) #3}
\newcommand{\prevd}[3]{Phys. Rev. {\bf D#1}, (#2) #3}
\newcommand{\nuclpb}[3]{Nucl. Phys. {\bf B#1}, (#2) #3}
\newcommand{\prog}[3]{Prog. Theor. Phys. {\bf #1}, (#2) #3}
\newcommand{\zeitc}[3]{Z. Phys. {\bf C#1}, (#2) #3}
\newcommand{\chp}{H^{+}}
\newcommand{\mhc}{m_{\chp}}
\newcommand{\mt}{m_{t}}
\newcommand{\mb}{m_{b}}
\newcommand{\mc}{m_{c}}
\newcommand{\ms}{m_{s}}
\newcommand{\mw}{m_{W}}
\newcommand{\mwl}{m_{W_1}}
\newcommand{\mwh}{m_{W_2}}
\newcommand{\df}{\frac{1}{(1-x)^4}}
\newcommand{\dt}{\frac{1}{(1-x)^3}}
\newcommand{\dty}{\frac{1}{(1-y)^3}}
\newcommand{\qt}{Q_t}
\newcommand{\bsg}{b \rightarrow s \gamma}
\newcommand{\bsemil}{b \rightarrow ce\bar{\nu}}
\newcommand{\bbsg}{Br(b \rightarrow s \gamma)}
\newcommand{\bbsemil}{Br(b \rightarrow ce\bar{\nu})}
\newcommand{\gbsg}{\Gamma(b \rightarrow s \gamma)}
\newcommand{\gbsemil}{\Gamma(b \rightarrow ce\bar{\nu})}
\newcommand{\cz}{c_\zeta}
\newcommand{\sz}{s_\zeta}
\newcommand{\cdb}{c_{2\beta}}
\newcommand{\sdb}{s_{2\beta}}

\begin{titlepage}
\rightline{~}
\rightline{BA-94-02}
\rightline{January, 1994}

\vspace*{1.5cm}

\begin{center}

\addtocounter{footnote}{1}

\centerline{\sc{{\Large \bf Supersymmetric Models With}}}
\vspace{0.5cm}
\centerline{\sc{ {\Large \bf
Tan${\bf \beta}$ Close to Unity}}}
\vspace{1.3cm}

{\sc B. Ananthanarayan$^{(1)}$, K.S. Babu$^{(2)}$ and Q. Shafi$^{(2)}$}
\vspace{3.5cm}

{\bf ABSTRACT}
\end{center}

Within the framework of supersymmetric grand unification, estimates
of the $b$ quark mass based on the asymptotic relation $m_b \simeq m_\tau$
single out the region with $\tan\beta$ close to unity, particularly if
$m_t(m_t) \stackrel{_<}{_\sim} 170\ GeV$. We
explore the radiative breaking of the electroweak symmetry and the associated
sparticle and higgs spectroscopy in models with
$1 < \tan\beta \stackrel{_<}{_\sim} 1.6$. The lightest scalar higgs is
expected to have a mass below
$100\ GeV$, while the remaining four higgs masses exceed $300\ GeV$.
The lower bounds on some of the sparticle masses are within the range
of LEP 200.
\vfill
7\noindent
\rule[.1in]{1in}{.01in}

\noindent
{\footnotesize $^{(1)}$ Institut de Physique Th\'{e}orique,
Universit\'{e} de Lausanne, CH 1015, Lausanne, Switzerland}\\
{\footnotesize $^{(2)}$ Bartol Research Institute, University of
Delaware, Newark, DE 19716, USA}

\end{titlepage}
\baselineskip = 0.7cm

\newpage

\dspace
Recent improvements in the determination of the three standard model
gauge couplings has renewed interest in the exciting possibility that
they do indeed converge to the same value at some superheavy scale
$\sim 10^{16}\ GeV$\cite{amaldi}. Within the framework of standard grand
unification (GUT), compatibility with the available data requires an
intermediate mass scale, on the order of $10^{12\pm 1}\ GeV$ in some
schemes. Perhaps a somewhat more elegant possibility is provided by
supersymmetric grand unification
(SUSY GUT), with SUSY broken at a scale $M_S \sim$ 100 GeV to a few TeV.
Among other things, this latter approach offers the prospects of
resolving the formidable gauge hierarchy problem\cite{dvali}.

One could also hope that SUSY GUTs can shed light on at least some of the
many parameters appearing in the minimal supersymmetric $SU(3) \times
SU(2) \times U(1)$ model (MSSM). A particularly important example is offered
by the well known asymptotic relation $m_b^0 \approx m_\tau^0$\cite{chanowitz}.
It has become
clear in recent years\cite{arason}
that the determination of the $b$-quark mass from this
relation depends in a rather interesting way on $m_t(m_t)$,
$\alpha_S(M_Z)$ and $\tan\beta$. In particular, and we verify this
in the first part of the paper, for $\alpha_S(M_Z)$ near 0.12 and
$m_t(m_t) \stackrel{_<}{_\sim} 170\ GeV$, the two regions, $\tan\beta$
close to unity $(1 \stackrel{_<}{_\sim} {\rm tan}\beta \stackrel{_<}
{_\sim} 1.6)$, and tan$\beta \gg 1$ are singled out. The large tan$\beta$
case can arise naturally in minimal $SO(10)$ type GUTs\cite{anant1}
and has been the
subject of several investigations\cite{anant2}.
However, weak-scale radiative corrections to
the $b$ mass through gluino exchange\cite{hall}
can be quite substantial in this case. With tan$\beta$ near
unity and a top mass in the range of 150 - 170 GeV (favored by
precision electroweak measurements), the top--quark Yukawa coupling
$h_t$ is near the infrared quasi--fixed point $h_t(m_t) \simeq 1.1$
\cite{pendleton}.
This may be an appealing feature in that a wide range of initial values
$h_t(M_X)$ near the GUT scale will be mapped to essentially the same
value of $h_t(m_t)$ thereby making it insensitive to the initial conditions.
Furthermore, with $\tan\beta$ close to unity, the weak-scale radiative
corrections to the $b$ mass can be quite small $(\stackrel{_<}{_\sim}
1-2 \%)$.

Some investigations have recently appeared in which $\tan\beta$ close
to unity has been considered, although not exclusively so
\cite{barger}. In this
paper we primarily focus on this possibility and, where appropriate,
offer comparisons with the large $\tan\beta$ case. It turns out that
the low $\tan\beta$ case is considerably less restrictive as far as
the allowed parameter space is concerned.  In particular, some of the
sparticles including a stau and stop may well be accessible to LEP 200.
With $m_t(m_t) \stackrel{_<}{_\sim} 170\ GeV$ the lightest CP even
scalar higgs mass is expected to be $\stackrel{_<}{_\sim} 100\ GeV$.

The starting point of our computations are the two-loop
renormalization group (RG) equations for the gauge and Yukawa couplings
\cite{barger2}. For
simplicity, we will only consider two possible values for the
(common) supersymmetry threshold $M_S$, a `low' value comparable to $m_t$,
and a `high value' of order a $TeV$.  Using $\alpha(M_Z) = 1/128$ and
sin$^2\theta_W(M_Z) = 0.233$ as inputs, we estimate the
unification scale $M_X$
to be about $2 \times 10^{16}\ GeV$, while the unified
gauge coupling $\alpha_G(M_X) \approx 1/26$.

The running $b$ quark mass has been estimated to be
$m_b(m_b) = 4.25 \pm 0.1\ GeV$
\cite{gasser}. Considering the various hadronic uncertainties
that are inherent in this determination, we will allow for a more generous
$2 \sigma$ error and take $m_b(m_b)$ to lie in the mass range $4.05 - 4.45\
GeV$.  The upper end will turn out to be crucial in
constraining tan$\beta$. It corresponds to a physical mass (pole mass) $M_b =
5.2\ GeV\ (5.1\ GeV)$ for $\alpha_s(M_Z) = 0.12\ (0.115)$.
A value of $M_b$ much larger than this seems unlikely.

The simplest (minimal) GUTs based on $SU(5)$ or $SO(10)$ predict the
asymptotic relation $m_b^0 = m_\tau^0$, which holds above the
unification scale $M_X \simeq 2 \times 10^{16}\ GeV$. Deviation from the
minimal GUT schemes, it appears, are necessary in order
to obtain realistic quark and lepton masses. Even then, the above
relation can hold to a good
approximation. For instance, implementation of the Fritzsch ansatz in
GUTs requires a non-minimal Higgs system
\cite{fritzsch}.  From the asymptotic
relation $Tr(M_\ell) = Tr(M_d)$,

\begin{equation}
m_d^0 - m_s^0 + m_b^0 = m_e^0 - m_\mu^0 + m_\tau^0
\end{equation}
Equation (6) leads to $m_b^0 = m_\tau^0 (1 - \epsilon)$, where
$\epsilon \approx (2/3)m_\mu^0/m_\tau^0 \approx 0.04$. Here $m_\mu^0
\simeq 3 m_s^0$ has been used.

Another possible source for fermion masses corresponds to
the non-renormalizable interactions with contributions of order

\begin{equation}
M_W \left( \frac{M_X}{M_{Pl}} \right)^n \stackrel{_<}{_\sim} {\cal O}
(100\ MeV),
\end{equation}

\noindent
with the $n=1$ contribution being of order the muon mass.  One expects the
relation $m_b^0 = m_\tau^0$ to hold at the 5\% level in this case.

Consequently, in addition to allowing a $2 \sigma$ error in
$m_b(m_b)$, we also allow for deviation of 5\% in the asymptotic
relation $m_b^0 = m_\tau^0$, i.e.,

\begin{equation}
m_b^0 = m_\tau^0 [1 \pm 0.05]
\end{equation}

In Fig. 1, we plot $m_b(m_b)$ versus $\tan\beta$ with $m_t = 150\
GeV$, $\alpha_s (M_Z) = 0.12$ and $M_S = m_t$. We have employed the
appropriate two-loop renormalization group equation for the running
of the gauge and Yukawa couplings between $M_X$ and $M_Z$, while
below $M_Z$ we have used the 3-loop QCD\cite{gorishny}
and one--loop QED
$\beta$ functions in the
evolution of $\alpha_s, \alpha$ and the mass parameters. The solid line in
Fig. 1 corresponds to $m_b^0 = m_\tau^0$, while the dot-dashed lines

account for the $\pm 5\%$ deviation from the exact equality. It can be seen
that there are two acceptable solutions
for $\tan\beta$, corresponding to a given value of $m_b(m_b)$. The
lower value is quite close to unity $(\tan\beta \simeq 1.1$ in this case),
while the larger values exceed $\tan\beta = m_t/m_b$. Barring radiative
corrections all
intermediate values of $\tan\beta$ are disfavored! If we require
$m_b(m_b) \stackrel{_<}{_\sim} 4.45\ GeV$, then $\tan\beta \stackrel{_<}{_\sim}
1.1$ or $\tan\beta \stackrel{_>}{_\sim} 60$ are singled out.

Of course, all this assumes that radiative corrections to the $b$ nd
$\tau$ masses can be ignored. It has been shown in Ref. [7] that
this very much depends on a combination of parameters which include
$\tan\beta$, the universal gaugino mass and the supersymmetric
higgsino mass. In the case that we are interested in, to wit,
$\tan\beta$ close to unity, we can a posteriori verify that the
radiative correction can be kept small ($\stackrel{<}{\sim}$
few \%), without imposing stringent new constraints on the available
parameter space.

The predictions for $m_b(m_b)$ are rather sensitive to the input values chosen
for
$\alpha_s(M_Z)$, $m_t$ and $M_S$. In Fig. 2, we plot the same curve,
but with a lower value of 0.11 for $\alpha_s(M_Z)$. Now the allowed
solutions are $\tan\beta \stackrel{_<}{_\sim} 1.1$ and $\tan\beta
\stackrel{_>}{_\sim} 42 (\approx m_t/m_b)$. In Fig. 3, $\alpha_s(M_Z) = 0.12$
is kept,
but $M_S = 1\ TeV$ is used. In this case, $\tan\beta \stackrel{_<}{_\sim}
1.1$ is the only allowed solution!

In Fig. 4, we display the variation of the prediction for $m_b(m_b)$
with the top-quark mass. Here, for clarity, $m_b^0 = m_\tau^0$ is the
only case displayed, with $m_t$ varying from $110\ GeV$ to $190\
GeV$. Also $\alpha_s(M_Z) = 0.12$ and $M_S = 1\ TeV$ have been used.
All intermediate values of $\tan\beta$ (say between
1.6 and 60) are excluded for $m_t \stackrel{_<}{_\sim} 170\ GeV$.
In Fig. 5 we have plotted an enlarged version of Fig. 4 near tan$\beta =
1$ for three values of $m_t$ (150, 160 and 170 GeV) from which it is clear
that tan$\beta \stackrel{_<}{_\sim}1.6$ at the lower end.
However, for $m_t \stackrel{_>}{_\sim} 180\ GeV$, all values of
$\tan\beta$  are allowed. This feature is displayed in Fig. 6 where
$m_t = 180\ GeV$ and $\alpha_s(M_Z) = 0.11$ are used.

Based on Figs. 1-6, the following conclusion can be reached. For $m_t
\stackrel{_<}{_\sim} 170\ GeV$ and $\alpha_s(M_Z) \stackrel{_>}{_\sim}
0.11$, and after allowing for uncertainties in the other relevant
quantities, the two regions $\tan\beta$ close to unity, or $\tan\beta
\gg 1$ are singled out.

It is an interesting
coincidence that the asymptotic mass relation $m_b^0 \simeq m_\tau^0$
prefers the two extreme values of tan$\beta$.  Since the parameter space
is narrow near the
end points, a certain amount of fine--tuning is needed to obtain the
desired values of tan$\beta$.  To see this explicitly, let us analyze
the tree--level (neutral) higgs potential of the MSSM:

\begin{equation}\begin{array}{rcl}
V_0 & = & \mu_1^2 |H_1^0|^2 + \mu_2^2 |H_2^0|^2 + \mu_3^2 (H_1^0
H_2^0 + h.c.)\strut\\
& + & \frac{g_1^2 + g_2^2}{8} (|H_1^0|^2 - |H_2^0|^2)^2\end{array}
\end{equation}

\noindent
where the mass parameters $\mu_i^2$ are defined as

\begin{equation}
\mu_1^2  =  m_{H_1}^2 + \mu^2\strut ;~~~
\mu_2^2  =  m_{H_2}^2 + \mu^2\strut;~~~
\mu_3^2  =  B\mu~~.
\end{equation}

\noindent
Here $\mu$ is the supersymmetric higgsino mass, and we let $v_1, v_2$
(both positive) denote the two vevs, such that $v^2 = v_1^2     +
v_2^2$ and $\tan\beta = v_2/v_1$.  Minimization of Eq. (4) yields
\begin{eqnarray}
{\rm sin}2\beta &=& {{2 \mu_{3}^2}\over {\mu_1^2+\mu_2^2}} \nonumber \\
v^2 &=& {2 \over {g_1^2+g_2^2}}\left[{{\mu_2^2-\mu_1^2}\over {{\rm cos}
2\beta}}-(\mu_1^2+\mu_2^2)\right]~~.
\end{eqnarray}
The large tan$\beta$ solution would require $\mu_3^2 \ll (\mu_1^2+\mu_2^2
)$.  The naturalness of this scenario has recently been discussed\cite{hall2}.

The scheme where tan$\beta$ is close to unity requires a different
fine--tuning condition.  Let us take the limit of strict equality,
tan$\beta=1$.  In this case,
\begin{equation}
\mu_1^2+\mu_2^2-2\mu_3^2=0
\end{equation}
from Eq. (6).
Furthermore, $\mu_1^2=\mu_2^2$ is needed to remedy the divergent
behavior in $v^2$ (see Eq. (6)).
In this limit, the Higgs potential has a custodial
$SU(2)$ symmetry (under which $H_1$ and $H_2$ rotate into each other).  This
symmetry, of course, is broken badly in the Yukawa sector where
$h_t \gg h_b$, as well as by the hypercharge gauge interactions.  These
effects show up in the Higgs sector only after one--loop
radiative corrections are taken into account.

The tan$\beta=1$ solution preserves the custodial $SU(2)$ symmetry (at
tree level), while the solution tan$\beta \gg 1$ breaks this
symmetry maximally.  In the former case, the Yukawa interactions
explicitly do not
respect the custodial symmetry, whereas in the latter case they do.
The hypercharge gauge interactions break this custodial symmetry
in either case.

As a consequence of Eq. (6), the $\tan\beta =1$ solution has a zero mass
scalar at tree--level which can be seen as follows.
The Higgs potential as a function of $v^2$, after minimizing with respect
to $\beta$, is given by
\begin{equation}
V(v^2)|_{{\rm tan}\beta=1} = v^2\left[{1 \over 2} (\mu_1^2+\mu_2^2)-
\mu_3^2 \right]~.
\end{equation}
The right-hand side of Eq. (8) vanishes as a consequence of Eq. (7),
leaving
$v^2$ undetermined.  The resulting ambiguity along the radial
direction gives rise to a (pseudo) Goldstone mode.
The relation (7) will be modified by the radiative corrections and
is to be regarded as the starting point of the SUSY analogue of the
Coleman-Weinberg mechanism\cite{diaz}. This has been studied in some detail in
the existing literature. The zero mass higgs mode, on account of the
large top quark Yukawa coupling, as
well as the existence of the scalar stops, can acquire a substantial
mass through radiative corrections. Within the MSSM, with the large
number of free parameters, the value of this mass can exceed $100\
GeV$. However, by embedding within a supergravity/GUT approach, we
find this to be not the case.

It is important to ask if $\tan\beta \approx 1$ can be realized in a
natural manner. One approach would be to replace the supersymmetric
higgsino mass term by $\lambda N(H_1 H_2-M^2)$, where $N$ denotes a
singlet superfield. The superpotential in this case possesses an
R-symmetry under which all matter superfields as well as the
superpotential change sign.  Now
the $SU(2) \times U(1)$ gauge symmetry can be broken
at tree level whilst preserving supersymmetry.  It is then ``natural''
to have tan$\beta \simeq 1$, since we can now take the limit where
all the SUSY breaking mass parameters are small compared to $M^2$.
Although intriguing, we will not
pursue this possibility any further here.

Our analysis of radiative electroweak breaking will be based on the
renormalization group improved tree level scalar potential given in Eq.
(4).  The various parameters evolve according to the well known one
loop RGE\cite{inoue} from $M_X$ to the low energy scale $Q_0$. A judicious
choice of $Q_0$\cite{gamberini} will yield a fairly reliable estimate of the
parameters, including bounds on the sparticle mass spectrum.

The spontaneous breaking of the electroweak gauge symmetry imposes
the constraint

\begin{equation}
\mu_1^2 \mu_2^2 \leq \mu_3^4
\end{equation}

\noindent
while the boundedness of the scalar potential yields

\begin{equation}
\mu_1^2 + \mu_2^2 \geq 2 |\mu_3|^2~~.
\end{equation}

The previous considerations of $m_b(m_b)$ help us pin down the
range of values at $M_X$ for the Yukawa couplings $h_b, h_\tau$ and
$h_t$. The parameter $\tan\beta$ at scale $Q_0 \sim M_S$ lies between 1 and
1.6. In addition, we have the parameters $M_\frac{1}{2},\
m_0,\ A$ at $M_X$, which denote the common gaugino mass, the common scalar
mass and the common trilinear scalar coupling. Once these are
specified, the one loop beta functions are integrated to yield the
gaugino and scalar mass parameters, as well as the trilinear
couplings at $Q_0$.

\noindent
Following Ref. \cite{gamberini}, we impose the constraint

\begin{equation}
|A(M_X)| < 3
\end{equation}

Note that the structure of the one loop beta functions does not
require us to know the values of the parameters $\mu$ (the SUSY
higgsino mass) and the bilinear coupling parameter $B (\equiv
\mu_3^2/\mu)$ at this stage
of the computation. Indeed, once $\tan\beta$ is known from the
considerations described earlier, we obtain from the minimization
conditions:

\begin{equation}
\mu^2(Q_0) = \left[ \frac{1}{2} M^2_Z(\tan^2 \beta-1) - (m_{H_1}^2 -
\tan^2 \beta m_{H_2}^2) \right] / (1 - \tan^2 \beta)~.
\end{equation}

\noindent
Further it can be seen that

\begin{equation}
\mu_3^2(Q_0) = \tan\beta \left[ - m_{H_2}^2 - \mu^2 - \frac{1}{2}
M_Z^2 \left( \frac{\tan^2\beta - 1}{\tan^2\beta + 1} \right) \right]~.
\end{equation}

There is a two fold ambiguity in the physical spectrum for a given
point in the parameter space due to the fact that $\mu$ can be of
either sign. This is found to affect the physical spectrum
significantly because of the structure of the terms in the sfermion,
neutralino and chargino mixing matrices when the electro-weak
symmetry is broken. We find that the dependence of the spectrum on
$A$ is minimal and therefore stick to the choice $A = 0$ for the
most part. The beta functions are such that the physical spectrum
remains unaltered under the interchange $(- M,A)$ with $(M, -A)$.
Thus, with $A = 0$ the physical spectrum shows no dependence on the
sign of $M$.

Armed with this a systematic search is performed to find the
acceptable region of the parameter space and the associated sparticle
spectrum. It may be seen that with all other parameters fixed, as
$\tan\beta$ approaches 1, the parameter $\mu$ rises. An unusually large
$|\mu|$ may be unattractive, and so we impose a somewhat ad hoc upper
bound of $1\ TeV$ on $|\mu|$
(radiative corrections to the b-mass can become significant
if $|\mu|$ is excessively large). This establishes a lower bound on $\tan\beta$
(separately for the choice of positive or negative $\mu$). For a
given of $\tan\beta$ there is a unique value of the common gaugino
and scalar mass for which the lightest neutralino is the LSP (albeit
degenerate with the lighter stau and stop). This follows from the
following observations. As $m_0$ is increased, with all other
parameters fixed, $\mu$ increases (thus an upper bound on $m_0$
emerges naturally). For a sufficiently small $\tan\beta$, as $m_0$ is
increased, the mass of the lighter scalar top falls, due to the
mixing term increasing in importance. However, a sufficiently large
$m_0$ is required to ensure that the lighter scalar tau is heavier
than the lightest neutralino (the would be LSP). As $m_0$ increases,
the bino-purity of the lightest neutralino is also found to increase.
We impose a constraint that the bino-purity of the lightest
neutralino be $\stackrel{_>}{_\sim} 95\%$\cite{olive} in order to ensure that
the
LSP is a good candidate for cold dark matter.

The above considerations are used in
determining the upper and lower bounds on the two primary soft SUSY
parameters $M_\frac{1}{2}$ and $m_0$. It is found that as $\tan\beta$
migrates away from its lower bound the parameter space opens up
somewhat. This is
because as $\tan\beta$ increases, even with fairly large values of
$M_\frac{1}{2}$ and $m_0$, we meet all of the consistency conditions,
with $|\mu|$ staying considerably smaller than a $TeV$.

We illustrate in the figures the behavior of the parameters of the
theory. In Figs. 7a, 7b we exhibit the upper and lower bounds on
$M_{\frac{1}{2}}$ (and the corresponding $m_0$ values) as functions
of $\tan\beta$. [Note that without the restriction $|\mu| \stackrel{_<}{_\sim}
1\ TeV$, the upper bound on $M_{\frac{1}{2}} \approx 800\ GeV$,
corresponding to a bino mass of $\approx 350\ GeV$.] Due to the fact that
for a given point in the parameter space the squark mixing is less
important when $\mu$ is negative, the lower bounds emerge smaller. In
Figs. 8a, 8b we show, for a particular choice of some of the parameters,
the variations of the lighter stau and stop masses as $m_0$ is varied.
In Fig. 9 we show the behavior for a larger value of $\tan\beta$. Comparing
Figs. 8 and 9 we find that the parameter space is much more constrained in the
former case. Finally, Figs. 10a, 10b display how the bino purity
varies with $m_0$, while Fig. 11 shows the variation of $|\mu|$ with
$m_0$ for fixed $M_{\frac{1}{2}}$.

To explain the shapes a brief discussion
is in order.  At the smallest value of $\tan\beta$ realizable,
we meet the conditions that $|\mu|\approx 1\ TeV$ and that the lighter
stop and stau are degenerate with the lightest neutralino.  As
$\tan\beta$ increases, we can find smaller values of the unified
gaugino mass such that the lighter stop and stau continue to be
degenerate with the lightest neutralino, with $|\mu|<1\ TeV$.
However, with increasing $\tan\beta$ the lightest neutralino begins to
mix strongly with
one of the higgsinos.
We require that the bino purity remain in excess of, and restrict the
parameter range further by requiring that the $<95\%$ lighter stau
mass be no larger than three times the LSP mass. The maximum value of
$M_{\frac{1}{2}}$ for a given $\tan\beta$ is reached when $m_0$ is
chosen such that the lightest neutralino is the LSP, and
simultaneously the upper bound on $|\mu|$ is saturated.
In Fig. 8, 9, the shapes of the stau and stop mass contours
are easily understood. Since the mixing term is
negligible for the stau, increasing $m_0$ leads to increasing
values of the stau. In the stop sector, however, as $m_0$ increases,
$|\mu|$ increases leading to appreciable mixing and the
diminishing of the smaller eigenvalue of the stop mixing
matrix, i.e., the mass of the lighter physical stop.
Indeed as discussed earlier, as $\tan\beta$ falls, this
mixing becomes even more pronounced for a given point in
the parameter space, and thus the parameter
space becomes more constrained since the stop must remain lighter than
the lightest neutralino.

It has been argued and found to be true by comparing our computations
with other related studies that the tree level potential is
consistent to within 10\% with the complete one loop scalar
potential. This is achieved by the selection of $Q_0$ to be of the
order of the geometric mean of the scalar top masses. We also study
the variations in the spectrum for $Q_0$ between the two
extreme limits of $M_Z$ and $1\ TeV$.

As far as the sparticles are concerned it is not inconceivable that
the lighter stau and stop will be found at LEP 200. Equally likely,
however, is the possibility that the sparticles are all heavy,
accessible perhaps only at the LHC. This is the case where the bino
LSP has mass $\sim 350\ GeV$, one of the staus is the lightest
charged sparticle, while the colored sparticles approach and even
exceed the $TeV$ range.

As far as the higgs sector is concerned, with $m_t(m_t)
\stackrel{_<}{_\sim} 170\ GeV,\ |\mu| <1\ TeV$, stop masses
$\stackrel{_<}{_\sim} 1-2\ TeV$, and $A_t(Q_0) \stackrel{_<}{_\sim}
1\ TeV$, the lightest (CP even) higgs mass is expected to be between
$60\ GeV$ and $100\ GeV$\cite{kodaira}.
It offers the best prospects for a truly remarkable
discovery at LEP 200. The CP odd scalar mass exceeds $300\ GeV$ which
implies that the remaining scalar higgs will not be accessible for
quite some time.

\bigskip

\noindent{\bf Acknowledgement:} B. A. thanks the Swiss National
Science Foundation for support during the course of this work.
K. S. B. and Q. S. thank the Department of Energy for partial
support during the course of this work under DOE Grant
DE-FG02-91ER406267.

\newpage

\noindent{\bf Figure Captions}

\noindent Fig. 1. Plot of $m_b(m_b)$ vs. $\tan \beta$ for $\alpha_S(M_Z)=0.12$.

\noindent Fig. 2. Plot of $m_b(m_b)$ vs. $\tan \beta$ for $\alpha_S(M_Z)=0.11$.

\noindent Fig. 3. Plot of $m_b(m_b)$ vs. $\tan \beta$ for a SUSY threshold of
1 TeV.

\noindent Fig. 4. Contours of $m_b(m_b)$ vs. $\tan \beta$ for $m_t(m_t)$
varying from 110 GeV to 190 GeV.

\noindent Fig. 5. Inset of Fig. 4 near $\tan \beta=1$.

\noindent Fig. 6. Plot of $m_b(m_b)$ vs. $\tan \beta$ for $m_t(m_t)=180$
GeV, $\alpha_S=0.11$.

\noindent Fig. 7a. Plots of mass parameters $M_{\frac{1}{2}}$ and $m_0$
vs. $\tan \beta$ for $\mu>0$.

\noindent Fig. 7b. Same as in Fig. 7a with $\mu < 0$.

\noindent Fig. 8a. Plot of bino, stop and stau mass vs. $m_0$
for a choice of parameters with $\mu > 0$.

\noindent Fig. 8b. Same as in Fig. 8a with $\mu < 0$.

\noindent Fig. 9a. Same as in Fig. 8a with a larger $\tan \beta$ value .

\noindent Fig. 9b. Same as in Fig. 8b with a larger $\tan \beta$ value.

\noindent Fig. 10a. Bino purity vs. $m_0$ for a choice of
parameters with $\mu > 0$.

\noindent Fig. 10b. Same as in Fig. 10a with $\mu < 0$.

\noindent Fig. 11. Plot of $|\mu|$ vs. $m_0$ for a choice of
parameters.

\end{document}